# Surface Reconstruction of Hexagonal Y-doped HoMnO$_3$ and LuMnO$_3$ studied using low-energy electron diffraction

R. Vasić[1], J. T. Sadowski[2], Y.J. Choi[3], H.D. Zhou[4], C.R. Wiebe[4], S.W. Cheong[3], J. E. Rowe[5], M.D. Ulrich[5,6]

[1] Department of Physics, Yeshiva University, New York, New York 10016, USA

[2] Center for Functional Nanomaterials, Brookhaven National Laboratory

   Upton, New York 11973, USA

[3] Department of Physics and Astronomy Rutgers, The State University of New Jersey

   Piscataway, New Jersey 08854, USA

[4] Condensed Matter Group/ Experimental, Florida State University, NHMFL FSU,

   Tallahassee, Florida 32310, USA

[5] Physics Department, NC State University, Raleigh, North Carolina 27695, USA

[6] Physics Division, Army Research Office, Research Triangle Park, North Carolina 27709, USA

*Abstract*

We have investigated the (0001) surfaces of several hexagonal manganite perovskites by low-energy electron diffraction (LEED) in order to determine if the surface periodicity is different from that of the bulk materials. These LEED studies were conducted using near-normal incidence geometry with a low energy electron microscope (LEEM)/LEED apparatus from room temperature to 1200°C and with an electron energy in the range of 15 – 50 eV. Diffraction patterns showed features of bulk-terminated periodicity as well as a 2×2 surface reconstruction. Possible origins for this surface reconstruction structure are discussed and comparisons are made with surface studies of other complex oxides.

## 1. Introduction

We report in this paper the surface periodic structure of the (0001) surfaces of several related hexagonal complex-oxide single crystals: $HoMnO_3$, $YMnO_3$, $LuMnO_3$, and h-$Ho_xY_{1-x}MnO_3$ (x=0.2, 0.6). The surface periodicity was observed by low energy electron diffraction (LEED) at room temperature and with heating from 500 to 1200°C, spanning several bulk structural phases. An empirical link between surface reconstruction and bulk phases would unambiguously indicate the importance of surface and interface structure in the functionality of complex oxides. Because our sample preparation and measuring geometry led to sufficient surface conductivity and/or charge neutralization from the LEEM electron gun, we did not observe the expected sample charging of these normally insulating materials. Typical LEED results for $HoMnO_3$, $YMnO_3$, and $LuMnO_3$ with electron energy of 30 eV are shown in Fig. 1 (a – c) with details discussed later in the results section.

The data in Fig. 1 show bulk-terminated surface LEED features of both the centrosymmetric and non-centrosymmetric phases and a 2×2 surface periodicity due to surface reconstruction. Surface reconstruction has not been universally observed in oxide surfaces (c.f. [1-4]). In complex multiferroic oxides, such surface reconstruction can stabilize ferroic order different from that of the bulk. For example $SrTiO_3$ is cubic but surface distortions induce ferroelectric order at the surface [5-7] and there is evidence of surface reconstruction. Hexagonal perovskite oxides such as $LiNbO_3$ show no evidence of surface reconstruction in LEED experiments [1-2]. The rare-earth hexagonal manganites *RMnO*$_3$ (*R* here represents from Ho to Lu, and Y) recently have attracted considerable attention due to the uncommon coexistence of coupled ferroelectric and antiferromagnetic ordering [8-16]. In bulk single crystal form these materials normally crystallize with the hexagonal structure, characteristic for rare earth ions of smaller radius (space group *P6$_3$cm*). While the Néel temperature is too low for some applications, these materials are of interest as model multiferroics for memory, electronic and spintronics applications. Optimizing multiferroic phenomena for technology will require an accurate understanding of the surfaces of multiferroics.

## 2. Experimental

All measurements were performed on single crystal samples of $R$MnO$_3$ ($R$ here represents Ho, Y and Lu). High quality single crystals of HoMnO$_3$, YMnO$_3$, and h-Ho$_x$Y$_{1-x}$MnO$_3$ (x=0, 0.2, 0.6, 1) were grown by the traveling solvent floating-zone (TFZ) technique at the National High Magnetic Field Laboratory (NHMFL). Single crystals of LuMnO$_3$ were grown by a flux technique at Rutgers University. Single TFZ crystals were oriented using Laue diffraction, cut and polished to yield (0001) surfaces. Polishing consisted of four steps during which a constant pressure of 50 k N/m$^2$ was applied. Polishing began with 30 µm SiC for 2 min, followed by 6 µm diamond for 2 min, 1 µm diamond for seven minutes and 20 nm blue colloidal silica for 2 min. The *rms* roughness of the samples over a 50 µm × 50 µm selected area after this process was 3.3 nm as measured by atomic force microscope (AFM). The samples were etched with a standard hydrofluoric acid solution before mounting for UHV experiments.

With additional measurements by AFM, a number of regions ~ 5 µm × 5 µm in size could be readily observed with a *rms* roughness of ~ 0.5 nm. Figure 2 shows typical data measured by AFM of the polished (0001) face of single crystal YMnO$_3$ at room temperature. The image size in this figure is 20 µm × 20 µm. The polishing step boundaries (tilted slightly from horizontal) are ~12 – 15 µm apart and are ~100 nm high. The flatter regions in between these steps have an *rms* roughness of ~2 nm. We estimate that the flatter regions are about 80-90% of the total surface area. The images were taken in contact mode and analyzed with the software program WSxM [17]; the surfaces of HoMnO$_3$, LuMnO$_3$ and Ho$_{.6}$Y$_{.4}$MnO$_3$ are similar to YMnO$_3$.

Low-energy electron diffraction studies were conducted in an Elmitec LEEM III system [18] at Brookhaven National Laboratory's Center for Functional nanomaterials [19]. Prior to LEED studies, the samples were degassed at 125°C for 2 hours. Low energy electron diffraction images were obtained with electron energies from 15 – 50 eV at several temperatures up to 1200°C for YMnO$_3$ and HoMnO$_3$, up to 1100°C for LuMnO$_3$, and up to 900°C for Ho$_{.6}$Y$_{.4}$MnO$_3$. Samples were heated in situ by radiation from a filament below 900°C and by electron bombardment for higher temperatures. The temperature was monitored by a thermocouple mounted on the sample cartridge. The temperature of the sample surface is estimated to be within ±20°C of this measurement.

In some cases, LEED patterns were obtained with electron energies up to 300 eV. However, the diffraction beams were typically too faint at large kinetic energies due to multiple scattering in the large $R$MnO$_3$ unit cell. LEED patterns were obtained in the μ-LEED mode, from regions 2 μm in diameter. This small sampling area allowed us to conduct measurements from regions with an *rms* roughness of less than 2.5 nm. After outgassing, but prior to heating, only the specular (00) LEED beam was observed. For sample temperatures of 500°C and above, diffraction beams were observed for all samples. For most samples the sharpness of the diffraction pattern improved with heating time up to 15-20 min; the best patterns being obtained after annealing samples at or above a temperature of 1100°C for 5 min and then cooling to ~300°C. At this temperature there may be sufficient ionic conductivity to prevent surface charging. The higher temperature of 1100°C likely introduced defects that reduced the sample resistivity and increased its electronic conductivity since LEED patterns always showed almost no evidence of charging after this anneal but for lower temperatures did display some charging effects which varied with the annealing history of each sample.

To obtain quantitative information about the in-plane lattice parameters in the investigated crystals, the diffraction patterns were referenced to data obtained from a well characterized Si(111)-7×7 surface, imaged with the same apparatus and instrumental settings. Calculations of real-space periodicities are based on first-order diffraction beams; for each sample, distances between (x,y) and (-x,-y) beams were measured for three sets of beams from patterns taken at each electron energy from 19.5 – 50 eV.

## 3. Results

As mentioned earlier, Fig. 1 shows the LEED patterns for HoMnO$_3$, YMnO$_3$, and LuMnO$_3$ taken with electron energy of 30 eV. Similarly sharp diffraction patterns were obtained at all energies from a low value of less than 10 eV up to a maximum of nearly 300 eV, although the intensity decreased with energy as expected. We were surprised to find that sharp diffraction patterns could be obtained at such low electron energies. It is reported that in an insulator, when the kinetic energy of the electrons is below ~50 eV, the incident electron flux exceeds the generation of secondary electrons. [20] The surface should charge negatively and reflect the incident electrons, preventing diffraction. We offer several potential explanations for the lack of charging which may be of interest for other materials when the conductivity is low.  While the LEEM/LEED apparatus utilizes a beam current which is more than 100 times lower than with conventional LEED display systems, the current density is higher. Unless the surface conductivity (which is unknown but could be higher) is much higher than the bulk conductivity, this would not explain the lack of charging.  Another important issue is the annealing to ~1100 or ~1200°C.  This may introduce near surface defects, which effectively dope the near surface region and thus increase the conductivity.  A final possible explanation is that the extreme demagnification of the LEEM/LEED beam may result in non-focused electrons that strike the sample holder and thus produce enough secondary electrons from the sample holder to neutralize the charge expected for these complex-oxide samples.

The LEED patterns of Fig. 1 are very similar for all three samples: $HoMnO_3$, $YMnO_3$, and $LuMnO_3$ with the assignments of these LEED beams shown in Figs. 1 (d) and 3.  The LEED patterns appear to indicate a simple hexagonal structure (see for example Fig. 1 (a) – 1 (c)), but a quantitative analysis reveals otherwise. The identification of periodicities revealed by the diffraction beams is based on mapping the calculated periodicity to the known crystal structure. All calculations were within 8% of the known structure. The accuracy of these measurements is influenced by sample alignment, aberrations in the electron optics and possibly by the disparity between semiconductor reference and oxide sample. The error would have to be at least 14% ($1-\sqrt{3}/2$) for the identification of the reconstruction to be ambiguous. All three compounds ($RMnO_3$: $R$=Ho, Y, or Lu) have a lattice parameter of ~6.1 Å in the non-centrosymmetric phase. In the centrosymmetric phase, the compounds have a (0001)-plane oxygen-oxygen (and equivalently $R$-$R$ and Mn-Mn) separation of 3.5 Å which in a triangular two-dimensional (2D) lattice gives a periodicity of 3.1 Å. This in-plane $R$-$R$ periodicity is preserved in the non-centrosymmetric phase which involves only a c-axis displacement of the R atoms. The displacement is 0.2 – 0.3 Å which is less than half the ionic radius of Y and the rare earth ions. Thus diffraction from rows of vertically displaced $R$ atoms is reasonable such that the (1,0) family of diffraction beams is evident in the non-centrosymmetric phase. We reference our 1×1 beams to this 3.1 Å periodicity and beams labeled 1 and 2 in Fig. 1(c) are identified accordingly. Note that this assignment is based on a quantitative analysis rather than the brightness of the beams. The (1,0) diffracton beams will not always be the brightest beams in the pattern as the brightness of each beam is a function of the electron kinetic energy. The beam labeled 1 (and the five equivalent beams at every 60°) are first order (1,0) diffraction beams corresponding to the 3.1 Å periodicity. Beams 2 (and the five equivalent beams at every 60°) are (1/√3,0)R30° beams corresponding to the real space periodicity of each of the $R$ atom sublattices in the non-centrosymmetric phase. In other words, they correspond to the first-order diffraction of the bulk-terminated (0001) 2D unit cell of the non-centrosymmetric phase. These are √3 times longer than in the centrosymmetric phase. A diagram of the $RMnO_3$ LEED pattern due to the bulk-terminated surface is shown in Fig. 1(d) with several LEED beams labeled according to this assignment scheme. Periodicities twice that of the 1×1 and (√3×√3)R30° are both observed in Fig. 1 evidencing a 2×2 reconstruction of the $RMnO_3$ basal plane. A diagram of the full $RMnO_3$ LEED pattern including the fractional beams is shown in Fig. 3 with several fractional beams labeled. The diffraction beams seen in Fig. 1(b) for $YMnO_3$ are elongated, suggesting additional structure. The elongation being along a principle direction is indicative of surface steps in which the elongation is related to the terrace width.

    To demonstration the quality of the diffraction studies, we show LEED-IV curves obtained with electron energies from 19.5 to 300 eV for the four beams, (1,0), (1/2,0), (1/√3,0)R30° and (1/2√3,0)R30° in Fig. 4. Only the (1,0) diffraction beam maintains a significant intensity through a kinetic energy of 300 eV. The intensity of the other beams becomes indistinguishable from noise above 100 eV. Because there are a large number of atoms in the unit cell, interference among multiple scattering events is more pronounced, causing the envelope of the LEED-IV curves to decrease rapidly. An analysis of this data is being conducted to determine the nature of the reconstruction.

    To further clarify the diffraction beam assignments, we discuss the crystal structure of the hexagonal manganites and the bulk-terminated surface. The bulk structure of $RMnO_3$ consists of layered vertex-sharing $MnO_5$ bipyramids separated by $R$ atoms. The R atoms are seven-fold coordinated in the non-centrosymmetric phase and 8-fold

coordinated in the centrosymmetric phase. The $Mn^{3+}$ and $R^{3+}$ ions have $MnO_5$ and $RO_8$ (centrosymmetric phase) local structures with bipyramidal $D_{3h}$ and trigonal $D_{3d}$ site symmetries, respectively [21]. In the non-centrosymmetric phase, the $MnO_5$ bipyramids are tilted toward or away from $R^{3+}$ ions and the rare earth layers are buckle such that there are two inequivalent $R^{3+}$ sites. Accordingly, the basal plane area of the non-centrosymmetric unit cell is three times that of the centrosymmetric unit cell. The bulk-terminated surface structures of the $RMnO_3$ basal plane for both the centrosymmetric (paraelectric) and the non-centrosymmetric (ferroelectric) phases are shown in Figs. 5 and 6 respectively. The surface unit cells of the observed surface periodicities in the non-centrosymmetric phase (shown in Fig. 6) are referenced to the (1×1) bulk-terminated surface unit cell of the centrosymmetric phase (Fig. 5). With this convention, the bulk-terminated surface of the non-centrosymmetric phase has a periodicity of $(\sqrt{3}\times\sqrt{3})R30°$. The surface unit cell corresponding to 2×2 periodicity is also shown in Fig. 6. This 2×2 periodicity is the surface reconstruction evidenced by the fractional beams in Fig. 1. This is not due to the bulk-terminated non-centrosymmetric surface. Here there are surface effects that are not yet understood.

Preliminary LEED studies of alloys of $Ho_xY_{1-x}MnO_3$ were also conducted and we found the same surface reconstruction as well as evidence of surface disorder. A typical LEED pattern, from the (0001) face of a single crystal of $Ho_{.6}Y_{.4}MnO_3$, is shown in Fig. 7. The sample was held at 900°C and observed at an electron energy of 25 eV. The labeling of beams corresponds to that of Fig. 1. The first-order $(1/2\sqrt{3},0)R30°$ diffraction beams are unresolved due to low intensity at this energy whereas the (1/2,0) diffraction beams are clearly evident. What appears to be higher order $(n/2\sqrt{3}, m/2\sqrt{3})R30°$ beams are faintly visible mostly below and to the right of the specular beam. However, these faint beams are due to multiple scattering.

The YMnO3, HoMnO3 and alloy samples demonstrated a broad transition from bulk termination to 2×2 surface reconstruction with increasing temperature. For $YMnO_3$ and HoMnO3, evidence of 2×2 surface reconstruction appeared at 750°C with sharp spots appearing at 1200°C. For $LuMnO_3$, the reconstruction was not evident at 900°C but was very clear at 1100°C. The alloy, $Ho_{.6}Y_{.4}MnO_3$, though not heated fully through the transition, gave evidence of reconstruction as low as 500°C. The presence of the family of (1/√3,0)R30° diffraction beams reveals the non-centrosymmetric phase. While present at lower temperatures, these beams were either missing or very faint at 1100 and 1200°C. The transition to the centrosymmetric phase for these oxides is slightly below this temperature and will depend upon growth and processing conditions. These results suggest that the vertical displacement of the *R* atoms at the surface follows that of the bulk transition. The 2×2 fully develops at these temperatures. Upon cooling, the (√3×√3)R30° periodicity returns and the presence of both periodicities dictates the presence of (2√3×2√3)R30°periodicity as is clearly revealed in Fig. 1.

## 4. Discussion

This study is the first systematic LEED study of several hexagonal transitional-metal/rare-earth manganite surfaces. Other diffraction studies of these materials has been conducted; primarily on YMnO$_3$. While several studies of hexagonal manganites have included electron diffraction in investigations of the bulk [22,-25], few have investigated the surface. In 1969, Aberdam *et al* observed 1×1 patterns in LEED studies of YMnO$_3$ [26] but did not observed the 2×2 reconstruction that we observe in all of our samples. The quality of samples and space charge effects are possible reasons [26].

The transition from a low-temperature ferroelectric to a high-temperature non-polar phase in hexagonal RMnO$_3$ with one intermediate phase is still a matter of debate [27]. From the current point of view, three phases are present in hexagonal manganites: triangular-ferroelectric (below Curie point ~625°C) and triangular-antiferroelectric (between Curie point and transition point to a nonpolar phase ~1080°C) both with space group *P6$_3$cm* and paraelectric above ~1080°C with space group *P6$_3$/mmc* [9,28]. The LEED studies reported here do not reveal in-plane structural changes primarily because the atomic positions in the basal plane shifts much less than the accuracy of the studies.

Care must be taken in the assignments discussed above because of the equivalence between first-order diffraction from one periodicity and higher order diffraction from another. For example, the first-order (1,0) diffraction beam associated with the bulk terminated centrosymmetric phase is coincident with the (2√3,1/√3)R30° diffraction beam. The assignment is based on the qualitative difference in the two families of diffraction beams; note the difference between the LEED-IV curves for the (1,0) and (1/√3,0)R30° beams shown in Fig. 4. Furthermore, diffraction from the (√3×√3)R30° periodicity disappears or is very faint at the highest temperatures while the 1×1 and 2×2 remain.

The comparison of the surface reconstruction reported here with results on other complex oxides is instructive. Matzdorf *et al* performed systematic study of surface-state spectroscopy in $Sr_2RuO_4$, a tetragonal oxide [29]. The layering in $Sr_2RuO_4$ is similar to h-$RMnO_3$ having corner sharing $RuO_6$ octahedra separated by Sr layers. In these studies they observed a surface reconstruction which is (√2×√2) R 45°. Their proposed mechanism for surface reconstruction is rotation of bipyramidal octahedral at the surface planes, but no distortion in the bulk planes. Another related compound is $LiNbO_3$, which is a ferroelectric with trigonal crystal symmetry. This material has also been studied by surface-science techniques reported by Yun *et al*. [1] Their 1×1 bulk terminated structure was clearly evident in all LEED patterns. The surface charge on this polar material originates from (un-ordered) oxygen adatoms and vacancies on Nb-terminated and Li-terminated surface respectively, but no reconstruction was distinguishable. Bharath *et al.* have also conducted LEED studies of $LiNbO_3$ without observing reconstruction [2]. In orthorhombic films such as $La_xSr_{1-x}MnO_3$ there is surface segregation of Sr atoms [3,30,31] but $La_xCa_{1-x}MnO_3$ does not segregate Ca atoms [31]. Neither reveals evidence of surface reconstruction. The correlation between surface reconstruction in layered oxides (the hexagonal manganites studied here, and $Sr_2RuO_4$) and the lack thereof in non-layered oxides ($LiNbO3$, and $La_xSr/Ca_{1-x}MnO_3$) suggests that perhaps layering decouples the surface from the bulk layers such that they are more amenable to reconstruction.

The origin of the surface reconstruction is unknown and will require further study. Several possibilities exist. From the ionic picture $RMnO_3$ the sheets are alternating $MnO_2^-$ and $RO^+$ layers. Theoretically, this would lead to a diverging electrostatic potential. In reality the surface charge is redistributed or the surface reconstructed [32]. Balancing the surface charge may be accomplished by vacancies or adatoms as in the $LiNbO_3$ case [1]. If so, the vacancies or adatoms order into a superstructure twice the size of the non-centrosymmetric unit cell. It is well known that the polar GaAs(111) surface has a 2×2 reconstruction due to ordering of a 1/4 ML of vacancies [33,34]. We are pursuing further studies to determine the nature of the reconstruction.

## 5. Conclusions

We have observed strong LEED patterns for $LuMnO_3$, $HoMnO_3$ and two Y-doped $HoMnO_3$ single crystals at room temperature and for temperatures from 500 to 1100°C. The diffraction patterns each have a periodicity in agreement with the lattice parameters of the bulk structure and also a 2×2 surface reconstruction. Ordering of oxygen vacancies or adatoms on the sample surface is suggested as a possible origin for the surface reconstruction. This might occur due to the driving force of reducing the polar nature of these surfaces in order to change the macroscopic dipole expected locally for a fully polar surface. These studies add to the growing body of results that indicate the need for more in-depth surface science studies of complex oxides in general.

**Acknowledgements:** We acknowledge Army Research Office for support for this research. Work at Rutgers was supported by DOE BES under Grant No. DE-FG02-07ER46382. The NHMFL is supported by contractual agreement between the National Science Foundation through NSF Grant No. DMR0449569 and the State of Florida. Research carried out in part at the Center for Functional Nanomaterials, Brookhaven National Laboratory, which is supported by the DOE BES under Contract No. DE-AC02-98CH10886. The National Synchrotron Light Source, Brookhaven National Laboratory, is supported by the DOE BES under Contract No. DE-AC02-98CH10886.

**Figure Captions:**

Fig. 1. LEED photographs of the (0001) face of single crystal (a) HoMnO$_3$, (b) YMnO$_3$ and (c) LuMnO$_3$ at 300°C observed after heating to > 1000°C. The electron energy is 30 eV. The LEED patterns indicate well-ordered surfaces with hexagonal symmetry. Part (d) gives the assignment discussed in the text. Two-dimensional Bragg beams given a (1,0) label correspond to the periodicity of rare-earth atoms in the centrosymmetric phase. A (1,0) diffraction rod is circled and labeled 1 in (c). Beams given a (1/√3,0)R30° label correspond to the bulk-terminated rare-earth sublattice periodicity of the non-centrosymmetric phase. A (1/√3,0)R30° diffraction rod is circled and labeled 2 in (c). For clarity, the other beams observed in the images are not included in (d).

Fig. 2. AFM topography of the polished (0001) face of single crystal YMnO$_3$ at room temperature; with image size: 20 μm × 20 μm. The polishing step boundaries (tilted slightly from horizontal) are ~12 – 15 μm apart and are ~100 nm high. The flatter regions in between these steps have an *rms* roughness of measured as ~2 nm. We estimate that the flatter regions are about 80-90% of the total surface area; the surfaces of LuMnO$_3$ and Ho$_{.6}$Y$_{.4}$MnO$_3$ are similar. A line profile is provided to illustrate the roughness in the flatter region. The images were taken in contact mode and analyzed with the software program WSxM. [17]

Fig. 3. A diagram of the observed diffraction pattern of the (0001) surface of h-*R*MnO$_3$ including the fractional beams. The large filled circles are the (1,0) beams. The large open circles are the (1/√3,0)R30° beams. The remaining beams are fractional beams.

Fig. 4. Representative LEED-IV curves extracted from diffraction data obtained with electron energies up to 300 eV from the polished (0001) face of single crystal HoMnO$_3$ at room temperature. Multiple scattering in a large unit cell dictates that the intensity of the three lower curves decreases rapidly with kinetic energy.

Fig. 5. Real space lattice of the bulk-terminated (0001) surface of h-*R*MnO$_3$ in the centrosymmetric phase. Large blue spheres are *R* atoms. The shaded triangles are the MnO$_5$ bipyramids and the small black dots are oxygen atoms. (a) Only the layer of MnO$_5$ bipyramids below the *R* atoms are shown. (b) The layer of MnO$_5$ bipyramids above the *R* atoms are shown to illustrate the stacking. The 1×1 unit cell is shaded.

Fig. 6. Real space lattice of the bulk-terminated (0001) surface of h-*R*MnO$_3$ in the non-centrosymmetric phase. The in-plane distortions are exaggerated for illustrative purposes. Large blue and medium red spheres are the *R* atoms in inequivalent sites. The large blue *R* atoms are displaced along the *c* axis (out of the page) approximately 0.2-0.3 Å from the plane of the medium red *R* atoms. The 1×1 and (√3×√3)R30° unit meshes (shaded) and 2×2 and (2√3×2√3)R30° unit meshes (unshaded) are shown.

Fig. 7. LEED image of the (0001) face of single crystal $Ho_{.6}Y_{.4}MnO_3$ held at 900°C and observed at an electron energy of 25 eV. The beams labeled 1 and 2 correspond to those in Fig. 1c. The $(1/\sqrt{3},0)R30°$ beams are not observable. Faint beams with distances from the specular beam intermediate between 1 and 2 are due to multiple scattering.

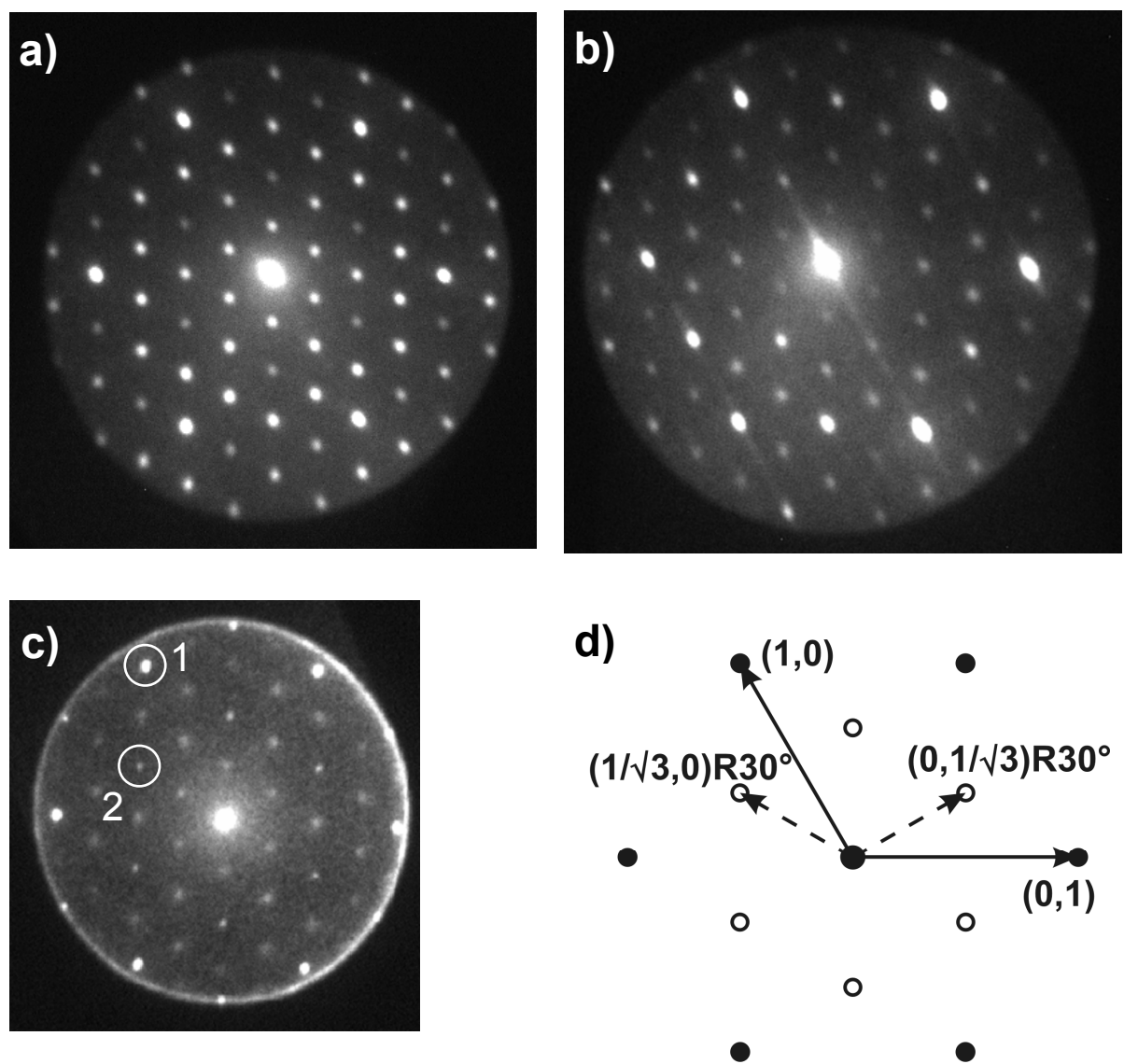

Fig. 1. LEED photographs of the (0001) face of single crystal (a) $HoMnO_3$, (b) $YMnO_3$ and (c) $LuMnO_3$ at 300°C observed after heating to > 1000°C. The electron energy is 30 eV. The LEED patterns indicate well-ordered surfaces with hexagonal symmetry. Part (d) gives the assignment discussed in the text. Two-dimensional Bragg beams given a (1,0) label correspond to the periodicity of rare-earth atoms in the centrosymmetric phase. A (1,0) diffraction rod is circled and labeled 1 in (c). Beams given a $(1/\sqrt{3},0)R30°$ label correspond to the bulk-terminated rare-earth sublattice periodicity of the non-centrosymmetric phase. A $(1/\sqrt{3},0)R30°$ diffraction rod is circled and labeled 2 in (c). For clarity, the other beams observed in the images are not included in (d).

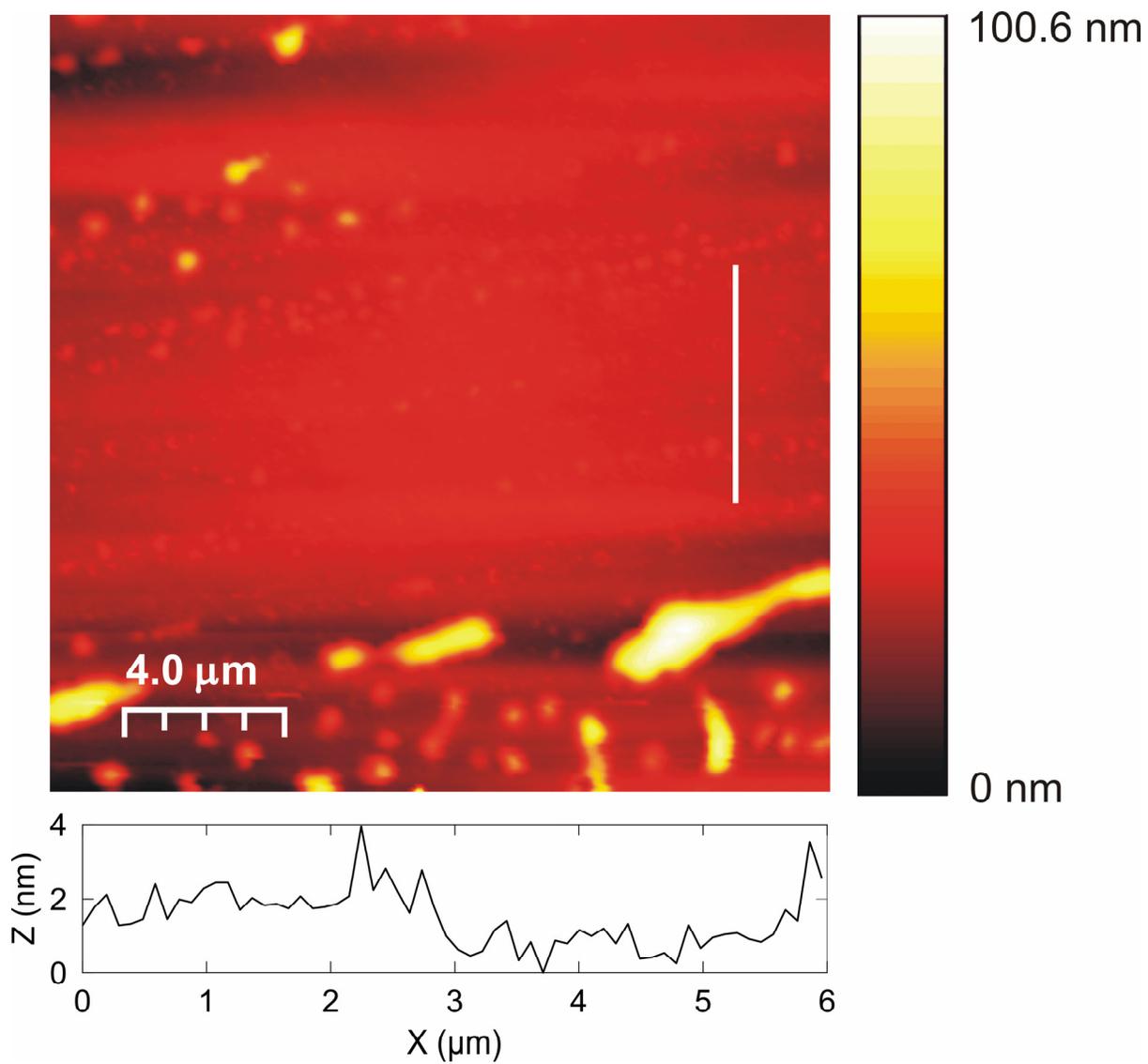

Fig. 2. AFM topography of the polished (0001) face of single crystal YMnO$_3$ at room temperature; with image size: 20 µm × 20 µm. The polishing step boundaries (tilted slightly from horizontal) are ~12 – 15 µm apart and are ~100 nm high. The flatter regions in between these steps have an *rms* roughness of measured as ~2 nm. We estimate that the flatter regions are about 80-90% of the total surface area; the surfaces of LuMnO$_3$ and Ho$_{.6}$Y$_{.4}$MnO$_3$ are similar. A line profile is provided to illustrate the roughness in the flatter region. The images were taken in contact mode and analyzed with the software program WSxM. [17]

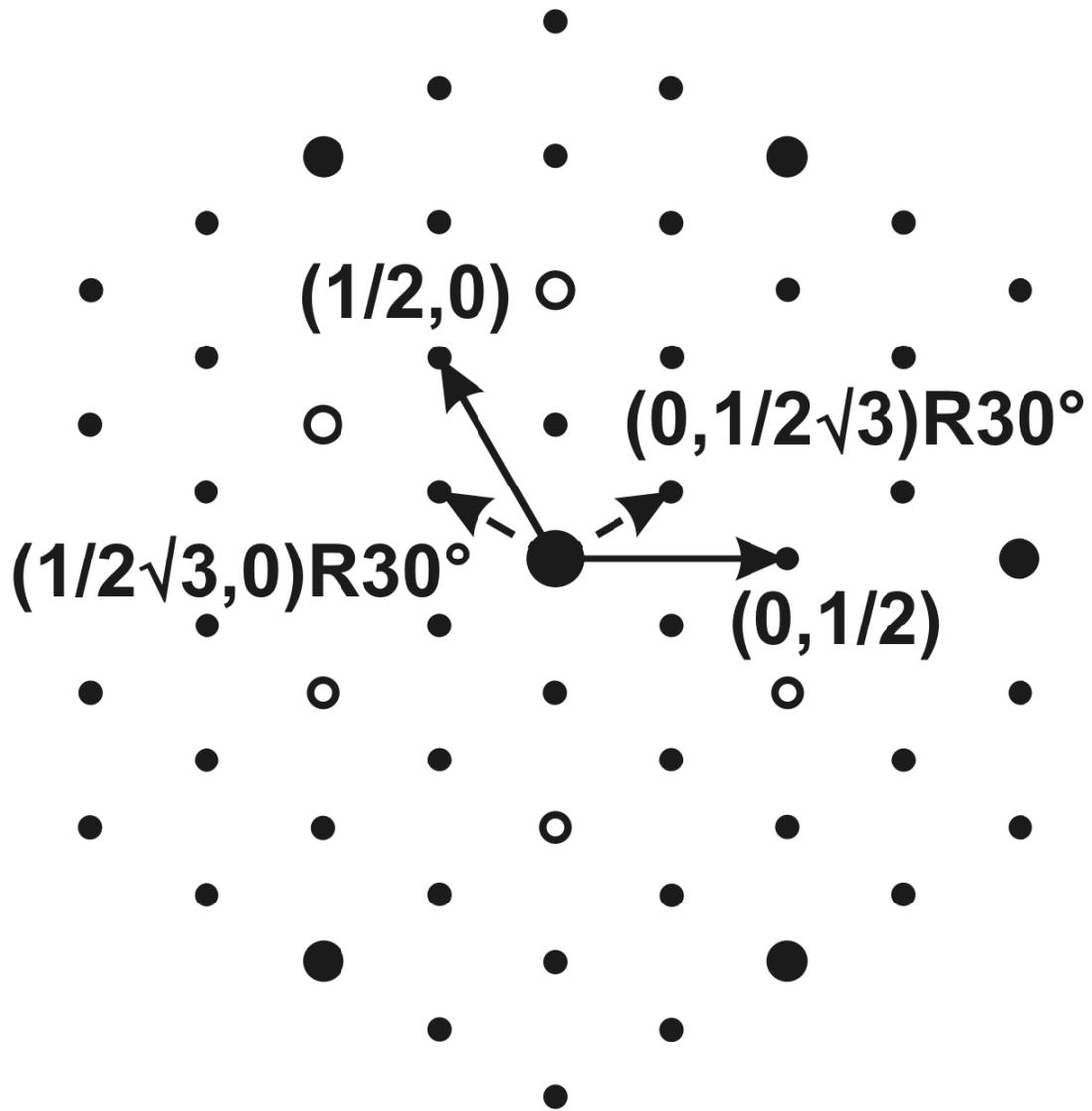

Fig. 3. A diagram of the observed diffraction pattern of the (0001) surface of h-RMnO$_3$ including the fractional beams. The large filled circles are the (1,0) beams. The large open circles are the (1/√3,0)R30° beams. The remaining beams are fractional beams.

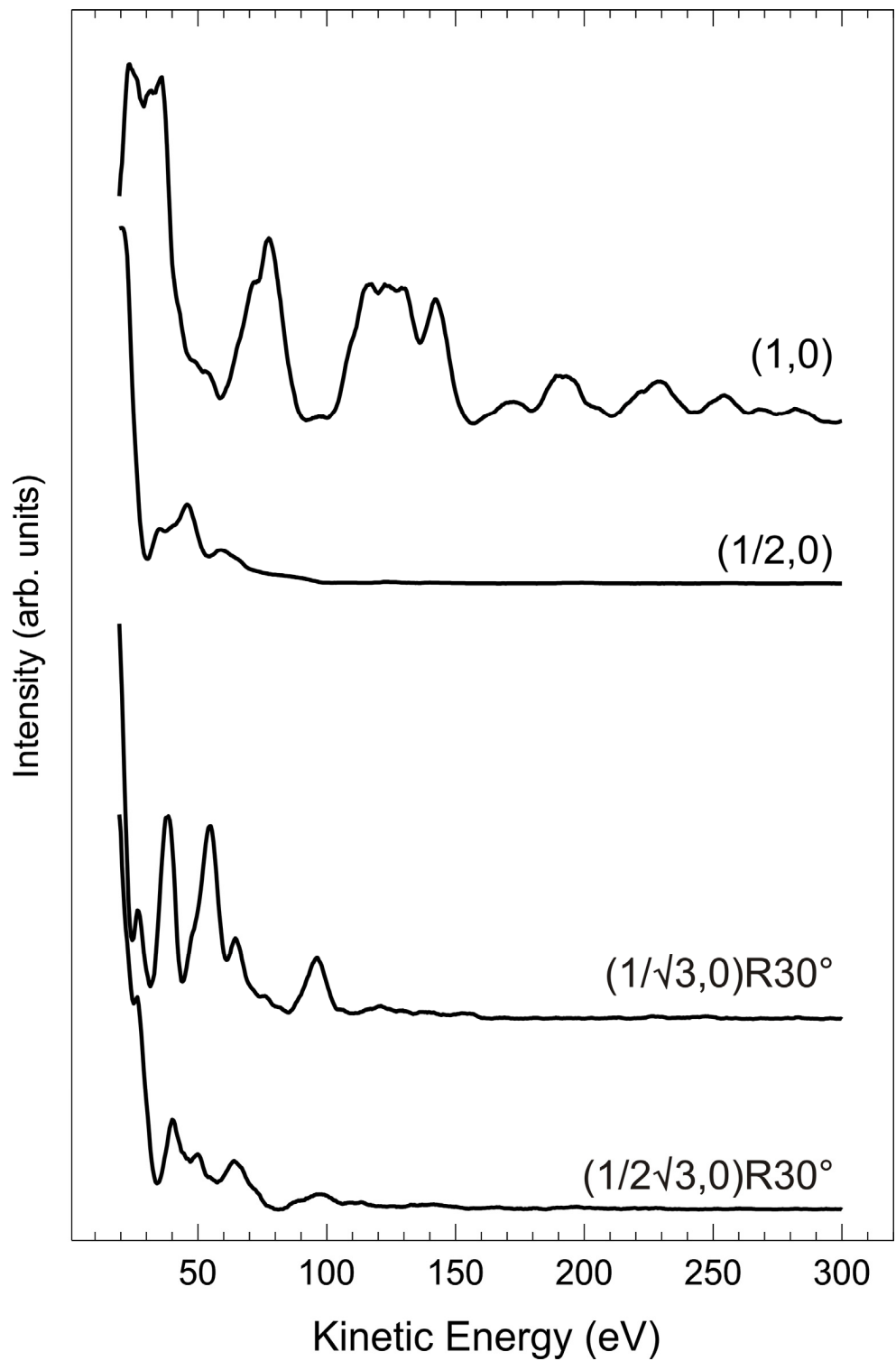

Fig. 4. Representative LEED-IV curves extracted from diffraction data obtained with electron energies up to 300 eV from the polished (0001) face of single crystal $HoMnO_3$ at room temperature. Multiple scattering in a large unit cell dictates that the intensity of the three lower curves decreases rapidly with kinetic energy.

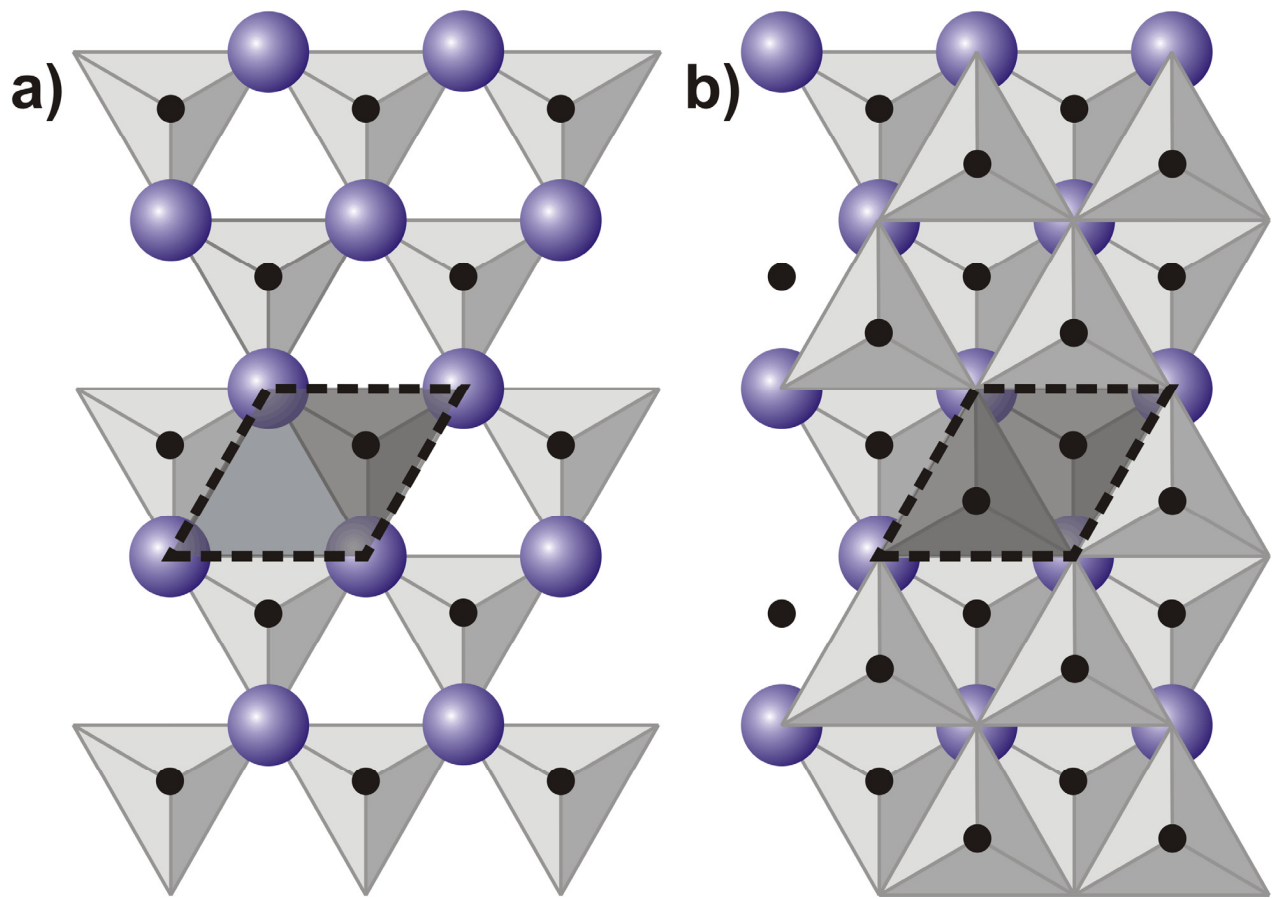

Fig. 5. Real space lattice of the bulk-terminated (0001) surface of h-$R$MnO$_3$ in the centrosymmetric phase. Large blue spheres are $R$ atoms. The shaded triangles are the MnO$_5$ bipyramids and the small black dots are oxygen atoms. (a) Only the layer of MnO$_5$ bipyramids below the $R$ atoms are shown. (b) The layer of MnO$_5$ bipyramids above the $R$ atoms are shown to illustrate the stacking. The 1×1 unit cell is shaded.

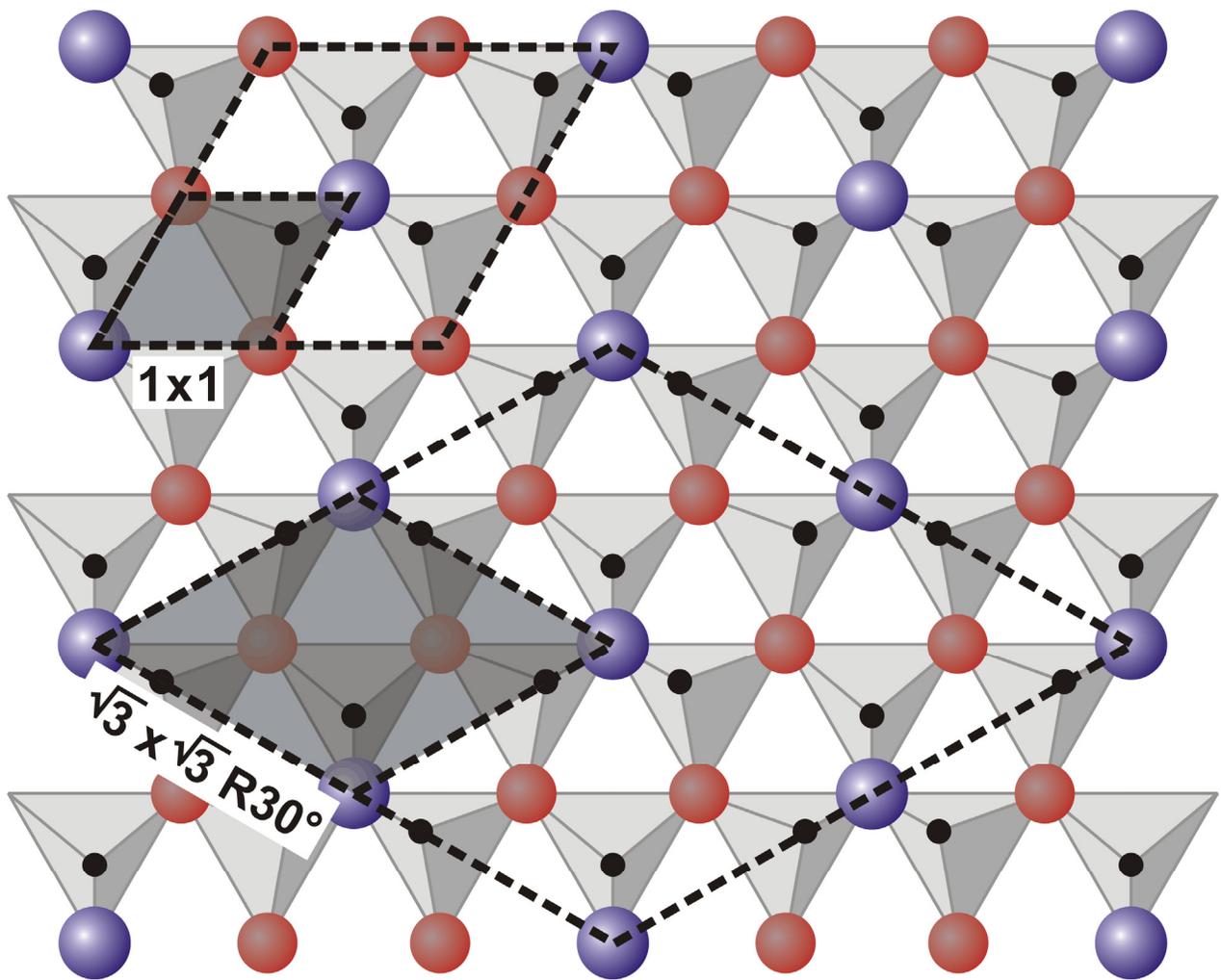

Fig. 6. Real space lattice of the bulk-terminated (0001) surface of h-RMnO$_3$ in the non-centrosymmetric phase. The in-plane distortions are exaggerated for illustrative purposes. Large blue and medium red spheres are the *R* atoms in inequivalent sites. The large blue *R* atoms are displaced along the *c* axis (out of the page) approximately 0.2-0.3 Å from the plane of the medium red *R* atoms. The 1×1 and (√3×√3)R30° unit meshes (shaded) and 2×2 and (2√3×2√3)R30° unit meshes (unshaded) are shown.

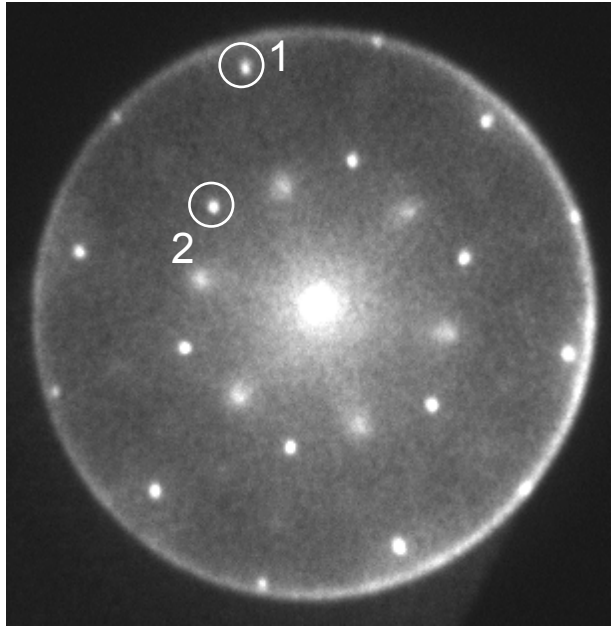

Fig. 7. LEED image of the (0001) face of single crystal $Ho_{.6}Y_{.4}MnO_3$ held at 900°C and observed at an electron energy of 25 eV. The beams labeled 1 and 2 correspond to those in Fig. 1c. The $(1/\sqrt{3},0)R30°$ beams are not observable. Faint beams with distances from the specular beam intermediate between 1 and 2 are due to multiple scattering.